\UseRawInputEncoding
\documentclass[12pt]{article}
\usepackage{epsfig,amsmath,amsfonts,amssymb,amstext,afterpage,psfrag,
}
\usepackage{slashed}

\usepackage[square,comma,sort&compress,numbers]{natbib}
\usepackage{dsfont}
\usepackage{framed}
\usepackage{placeins}
{\endMakeFramed}
\allowdisplaybreaks
\newdimen\TW
\begin{document}

\begin{titlepage}

\begin{flushright}
{\small
LMU-ASC~02/21\\
February 2021\\
}
\end{flushright}

\vspace{0.5cm}
\begin{center}
{\Large\bf \boldmath
Top-pair production via gluon fusion in the \\Standard Model Effective Field Theory\\
\unboldmath}
\end{center}

\vspace*{0.4cm}

\begin{center}
{\sc  Christoph M\"uller\footnote{E-mail: Christoph.Mueller1@physik.uni-muenchen.de}}
\end{center}

\vspace*{0.4cm}

\begin{center}
Ludwig-Maximilians-Universit\"at M\"unchen, Fakult\"at f\"ur Physik,\\
Arnold Sommerfeld Center for Theoretical Physics,
D-80333 M\"unchen, Germany\\
\end{center}

\vspace{1.5cm}
\begin{abstract}
\vspace{0.2cm}\noindent
We compute the leading corrections to the differential cross section for top-pair production via gluon fusion due to dimension-six operators at leading order in QCD. The Standard Model fields are assumed to couple only weakly to the hypothetical new sector. A systematic approach then suggests treating single insertions of the operator class containing gluon field strength tensors on the same footing as expli\-citly loop suppressed contributions from four-fermion operators. This is in particular the case for the chromomagnetic operator $Q_{(uG)}$ and the purely bosonic operators $Q_{(G)}$ and $Q_{(\varphi G)}$. All leading order dimension-six contributions are consequently suppressed with a loop factor $1/16\pi^2$.

\end{abstract}

\vfill

\end{titlepage}

\section{Introduction}
After the discovery of the Higgs-boson in 2012, a lot of attention has been paid to the unknown physics beyond the Standard Model (SM). While at proton colliders like the Large Hadron Collider (LHC) focus has mostly been put on the direct detection of new resonances by virtue of high center-of-mass energies, searching for indirect evidence of new particles through their virtual effects on the interactions between SM particles might represent a fruitful alternative. The latter approach has the advantage of being able to probe regimes that are beyond the kinematic energy bounds of the LHC. As the physics beyond the SM is a priori unknown, one has to employ a bottom-up effective field theory in order to systematically parameterize the new physics. A model independent approach with as few assumptions about the new physics sector as possible is provided by the Standard Model Effective Field Theory (SMEFT), which essentially enlarges the at most four-dimensional operators of the SM with non-renormalizable higher dimensional ones. A complete basis of up to operator dimension six has been given in \cite{Buchmuller:1985jz,Grzadkowski:2010es} and is commonly referred to as the "Warsaw basis". Recent developments have included complete sets of dimension eight and nine \cite{Murphy:2020rsh,Li:2020xlh}. However, within the framework of a bottom-up effective field theory further assumptions about the underlying nature of the new and modified interactions still have to be made. In this context several questions should be addressed, in particular the question whether higher loop orders have to be included for a given fixed canonical order calculation. This crucially depends on the coupling strengths of the SM fields to the new sector. For top-quark decay it has already been noted that higher loop order corrections might be important \cite{Boughezal:2019xpp}, in which case the respective running of the SMEFT operators should be taken into account. For instance, the renormalization group equations for magnetic-moment-type operators \cite{Jenkins:2013wua} suggest that a cancellation of the unphysical renormalization scale can only be achieved upon adding the correct four fermion-operators to the analysis.\\
This work about gluon fusion top-pair production serves as an example of how to treat the operator class containing gluon field strength tensors in general with the only assumption about the new physics sector being its weak coupling to the SM. Higher loop order corrections associated with four-fermion operators are in fact necessary to obtain consistent results.\\
As we intend to focus on the main calculational aspects, we have made some phenomenological simplifications for the sake of clarity. First, since cross sections in QCD factorize into soft and hard parts of the interaction, we may focus on the pure parton level reaction. In reality, parton distribution functions for the initial hadrons as well as jets associated with the final quark pair play crucial roles for experimental constraints that should be taken into account for a full phenomenological analysis. Second, we concentrate on the gluon fusion channels, since for instance at the LHC quark fusion becomes less important than gluon fusion with increasing energy. An in-depth review for top-pair production is given in \cite{Baernreuther:2012}. All contributions of the tree-level operators for gluon fusion top-pair production in SMEFT together with a comment on the assumptions about the underlying theory can be found in \cite{Zhang:2010dr}.\\
This paper is organized as follows: The general notational setup together with a summary of the SM calculations for gluon fusion top-pair production is done in section \ref{overview}. In section \ref{calculation} we identify the relevant SMEFT operators and their respective contributions. We also discuss how to treat the chiral projectors and comment on the renormalization program. A brief phenomenological study of some of the new contributions is given in section \ref{numerics}. All our analytical results for the cross sections are listed in an appendix.
\section{Overview and Standard Model result}\label{overview}
\begin{figure}
	\centering
	\includegraphics[width=0.4\textwidth]{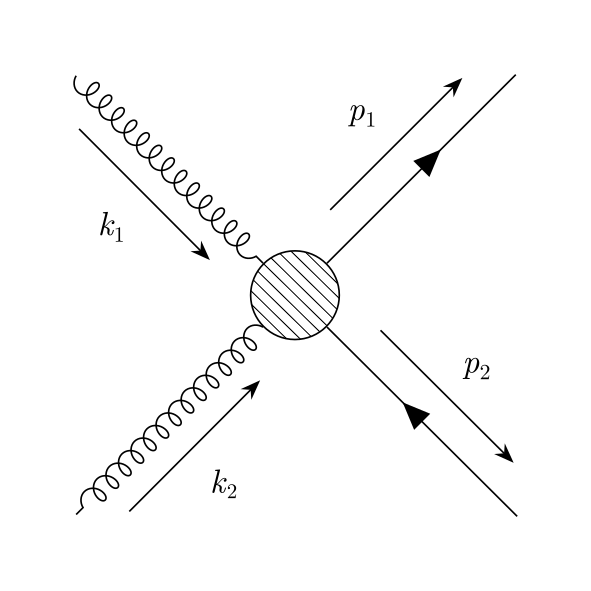}
	\caption{\textit{Kinematic setup for gluon fusion top-pair production with incoming momenta $k_1$ and $k_2$ and outgoing momenta $p_1$ and $p_2$. The circle in the middle represents all possible interactions for the theory under consideration. There are three Feynman diagrams within the SM at leading order. The SMEFT diagrams will be discussed below.}}
	\label{fig:fig}
\end{figure}

We consider the parton level process $gg\longrightarrow  t\bar t$ where two initial gluons merge to produce a top- and an antitop quark (see Figure \ref{fig:fig}). The unpolarized differential cross section in the center-of-mass frame in terms of the amplitude $\mathcal{M}$ is then given by the formula
\begin{align}
\frac{d\sigma}{d\Omega}=\frac{\overline{|\mathcal{M}|^2}}{64\pi^2s}\frac{|\boldsymbol{p_1}|}{|\boldsymbol{k_1}|}
\end{align}
where in $\overline{|\mathcal{M}|^2}$ an average over initial and a sum over final spins and colors is understood and $|\boldsymbol{p_1}|$ and $|\boldsymbol{k_1}|$ denote the spatial momenta of the outgoing top-quark and either incoming gluon. It can generally be written as a function of the relevant coupling constants, the Mandelstam variables $s$, $t$ and $u$ and the masses of the involved particles. The SM result is straightforwardly obtained by evaluating three diagrams and is given by \cite{Zyla:2020zbs}
 \begin{align}
\left(\frac{d\sigma}{d\Omega}\right)_{SM}=\frac{\alpha_s^2}{32s}&\sqrt{1-\frac{4m_t^2}{s}}\frac{7 m_t^4-7 m_t^2 (t+u)+4 t^2-t u+4 u^2}{3 s^2 (m_t^2-t)^2 (m_t^2-u)^2} \Bigl(t u (t^2+u^2)+\nonumber\\
&-6 m_t^8+m_t^4 (3 t^2+14 t u+3 u^2)-m_t^2(t + u) (t^2 + 6 t u + u^2)\Bigr)
\end{align}
where $\alpha_s=g_s^2/4\pi$ is the strong fine structure constant and $m_t$ is the top-mass.
\section{SMEFT calculation}\label{calculation}
The first contributions to $gg\longrightarrow \bar t t$ from operators of mass (canonical) dimension greater than four appear at dimension six and are accordingly suppressed by $1/\Lambda^2$, where $\Lambda$ denotes a potentially large scale of new physics. In our calculation we consider the interference terms of these new operators with the SM amplitude
\begin{align}\label{eq3}
|\mathcal{M}|^2=|\mathcal{M}_0|^2+\frac{1}{\Lambda^2}\left(\mathcal{M}_0^*\mathcal{M}_1+\mathcal{M}_1^*\mathcal{M}_0\right)+\mathcal{O}\left(1/\Lambda^4\right)
\end{align}
where the indices $0$ and $1$ refer to the SM and SMEFT amplitudes, respectively. Within this notation the canonical dimension of the various terms is manifest. However, it is important to notice that for effective theories a more general power counting prescription needs to be specified, even if it is linearly realized \cite{Buchalla:2013rka,Buchalla:2013eza,Buchalla:2016bse}. Before we dive into the actual calculation, we review some aspects of power counting in SMEFT.
\subsection{Remarks on power counting in SMEFT}
Higher dimensional operators are naturally organized in terms of their canonical dimension that indicates their relative suppression with respect to the new physics scale $\Lambda$ (see Equation (\ref{eq3}) above). This corresponds to an expansion in the dimensionless parameter $E/\Lambda$, where $E$ is a typical energy scale of the process, and from the point of view of a renormalizable field theory, the canonical dimension remains the strongest tool for a systematic approach. However, canonical dimensions alone do not provide enough information for a consistent treatment of the new operators. In addition, one needs to specify the loop order of a specific operator within the relevant process by including the expansion parameter $1/16\pi^2$ to the analysis. The necessity of keeping track of loop orders has been put on solid grounds in \cite{Arzt:1994gp} and can be discussed quantitatively in terms of chiral dimensions in a more general manner, see \cite{Buchalla:2013eza,Buchalla:2016sop}. For instance, in \cite{Arzt:1994gp} it has been argued that on the one hand, operators containing field strength tensors cannot be generated at tree-level when the underlying theory is assumed to be a weakly coupled gauge theory. In fact, the matching procedure has to occur at the one-loop-level at least. This has also been noted in \cite{Grzadkowski:2010es}. On the other hand, four-fermion operators can in fact be generated at tree-level.\\ The arguments presented in \cite{Arzt:1994gp} remain true for any underlying theory as long as it couples only weakly to the SM fields, which is the case for a wide class of possible high energy scenarios. The respective Wilson coefficients of operators containing field strength tensors should consequently be suppressed by the loop factor $1/16\pi^2$ when compared to the ones of the four-fermion operators in this case. Likewise it is easy to construct model theories that contradict the na\"ive assignment of an $\mathcal{O}(1)$-number to every Wilson coefficient. When a model independent parameterization like SMEFT is chosen, it therefore seems natural to stick to a weakly coupling new sector and to adopt the corresponding rules for a superficial estimation of the Wilson coefficients. Applying them to gluon fusion top-pair production, we find that there does not exist a pure tree-level contribution within the SMEFT at leading order in QCD at all. On the contrary, the first non-vanishing corrections are suppressed with respect to both $1/\Lambda^2$ \textit{and} $1/16\pi^2$, i.e. they are all one-loop suppressed. We will provide a complete list of the relevant operators and their particular contributions in the next section.
\begin{figure}
	\centering
	\includegraphics[width=0.9\textwidth]{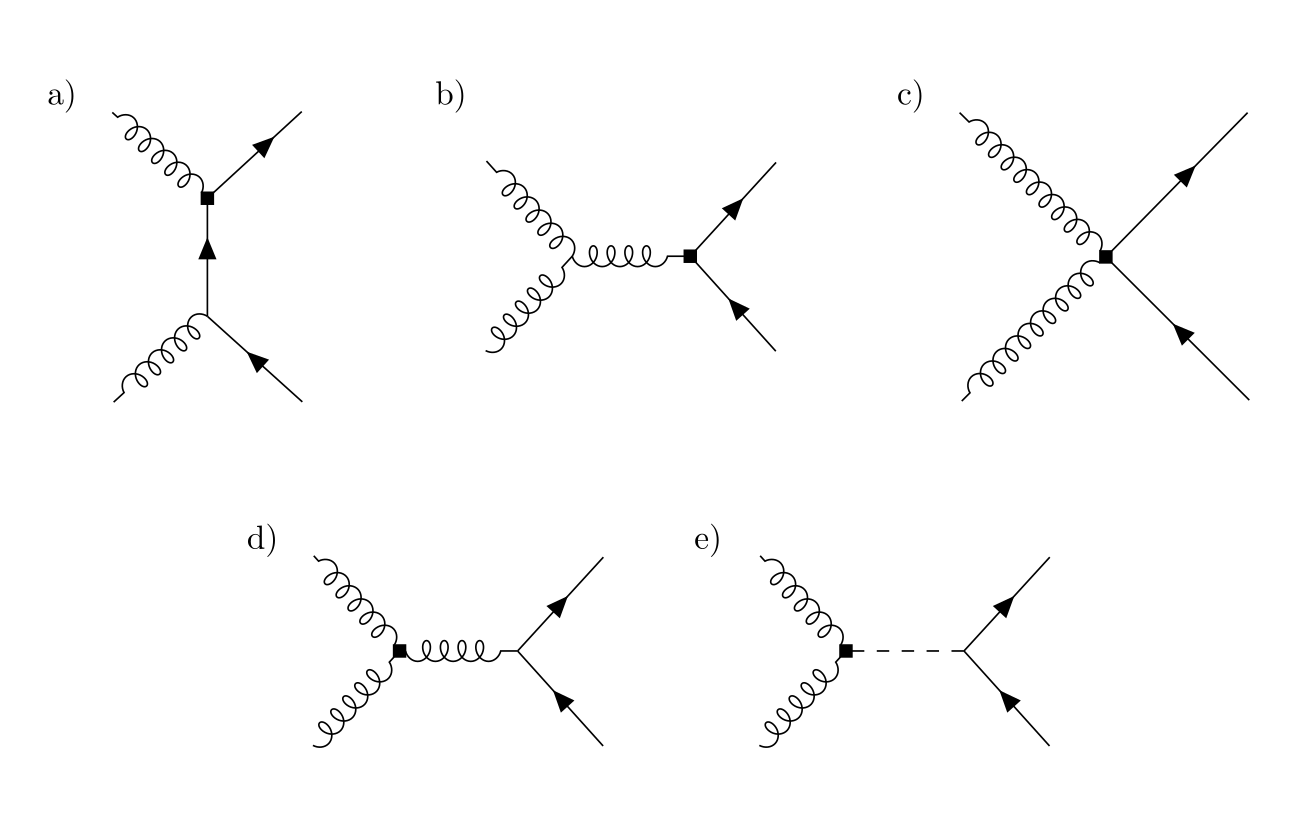}
	\caption{\textit{Feynman diagrams for the tree contributions in SMEFT, where dimension-six insertions are denoted by black squares. Crossings are not displayed.} a - c) \textit{Contributions from $Q_{(u G)}$. The local interaction} c) \textit{does not appear in the SM.} d) \textit{Contribution from $Q_{(G)}$ which modifies the s-channel process of the SM.} e) \textit{New contribution from $Q_{(\varphi G)}$.}}
	\label{fig:fig2}
\end{figure}
\FloatBarrier
\subsection{Relevant operators}\label{cal}
\begin{figure}
	\centering
	\includegraphics[width=0.9\textwidth]{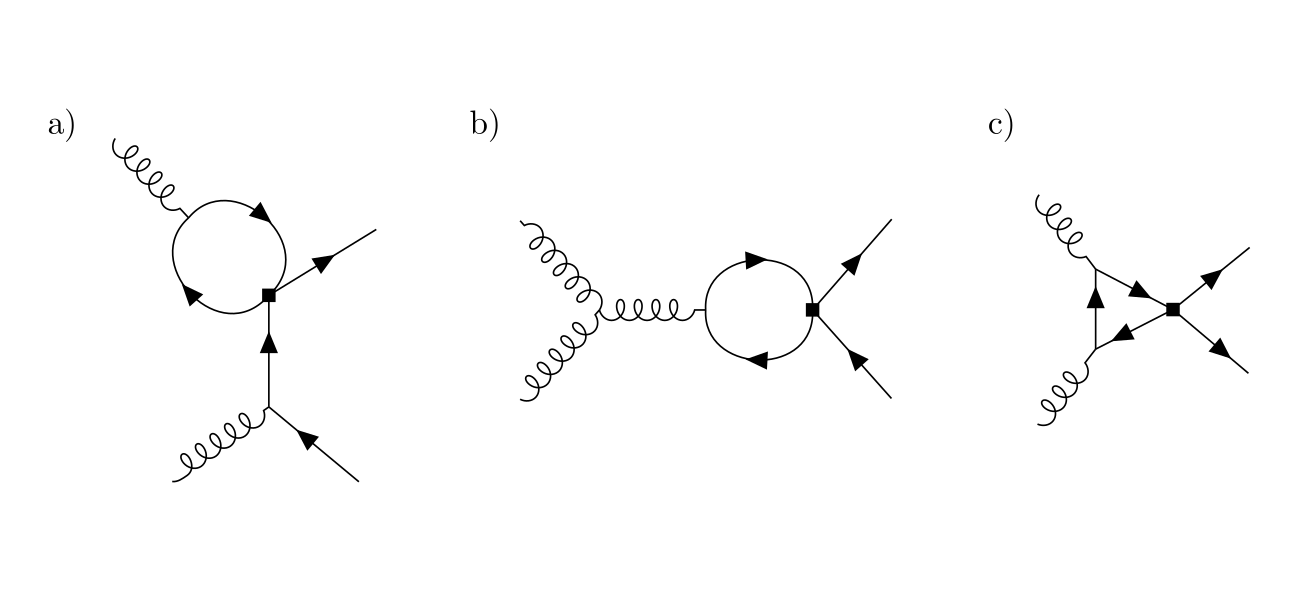}
	\caption{\textit{Feynman diagrams for the one-loop contributions in SMEFT. Again, crossings are not displayed. These diagrams are needed to cancel the implicit dependence on the renormalization scale $\mu$ in the diagrams of} Figure \ref{fig:fig2} a - c).}
	\label{fig:fig3}
\end{figure}

As the new physics is expected to mainly affect the third particle generation, we restrict our calculation to this sector and neglect effects of the CKM-matrix. We also ignore CP-odd operators and impose baryon number conservation. In the Warsaw basis the following operators are relevant for gluon fusion top-pair production at tree-level
\begin{alignat}{3}
&Q_{(G)}&&=&& \ f^{ABC}G_\mu^{A\nu}G_\nu^{B\rho}G_\rho^{C\mu}\nonumber\\
&Q_{(\varphi G)}&&=&& \ \varphi^\dagger\varphi G_{\mu\nu}^AG^{A\mu\nu}\nonumber\\
&Q_{(u G)}&&=&& \ (\bar q \sigma_{\mu\nu}T^A t) \tilde\varphi G^{A\mu\nu}	\label{treeoperators}
\end{alignat}
where $q$ is the left-handed third generation quark-doublet, $t$ is the right-handed top-quark, $\varphi$ is the Higgs-doublet ($\tilde \varphi^i=\epsilon^{ij}\varphi^{*j}$) and $G_{\mu\nu}^A$ is the $A$th component of the gluon field strength tensor with respect to the $SU(3)$ generators $T^A=\lambda^A/2$, where $\lambda^A$ are the Gell-Mann matrices. We adopt the conventions found in \cite{Dedes:2017zog}. These operators lead to the diagrams displayed in Figure \ref{fig:fig2}. Although their contributions are eventually loop suppressed by virtue of their Wilson-coefficients, we will refer to the operators in Equation (\ref{treeoperators}) as the "tree-level operators". Note that the local interaction between gluons and the Higgs-particle introduces a new s-channel contribution associated with the Higgs-mass $m_h$. Also, the chromomagnetic operator $Q_{(u G)}$ induces a new local interaction between two gluons and two quarks. Its role for gluon fusion Higgs production to higher loop orders has been investigated in \cite{Franzosi:2015osa,Deutschmann:2017qum}. For all tree-level operators we find full agreement with the analytic expressions for the differential cross sections of gluon fusion top-pair production found in \cite{Brivio:2019ius,Zhang:2010dr}.\\
In view of the last section the tree-level operators have to be supplemented by the following four-fermion operators
\begin{alignat}{6}
&Q_{(qd)}^{(1)}&&=&& \ (\bar q \gamma_\mu q)(\bar b \gamma^\mu b)\phantom{AAAAAAA}
&&Q_{(qd)}^{(8)}&&=&& \ (\bar q \gamma_\mu T^A q)(\bar b \gamma^\mu T^A b)\nonumber\\
&Q_{(ud)}^{(1)}&&=&& \ (\bar t \gamma_\mu t)(\bar b \gamma^\mu b)\phantom{AAAAAAAa}
&&Q_{(ud)}^{(8)}&&=&& \ (\bar t \gamma_\mu T^A t)(\bar b \gamma^\mu T^A b)\nonumber\\
&Q_{(qu)}^{(1)}&&=&& \ (\bar q \gamma_\mu q)(\bar t \gamma^\mu t)\phantom{AAAAAAA}
&&Q_{(qu)}^{(8)}&&=&& \ (\bar q \gamma_\mu T^A q)(\bar t \gamma^\mu T^A t)\nonumber\\
&Q_{(qq)}^{(1)}&&=&& \ (\bar q \gamma_\mu q)(\bar q \gamma^\mu q)\phantom{AAAAAAA}
&&Q_{(qq)}^{(3)}&&=&& \ (\bar q \gamma_\mu \tau^I q)(\bar q \gamma^\mu \tau^Iq)\nonumber\\
&Q_{(uu)}&&=&& \ (\bar t \gamma_\mu t)(\bar t \gamma^\mu t)\phantom{AAAAAAA}
&&Q_{(quqd)}^{(1)}&&=&& \ (\bar q^j t)\epsilon_{jk}(\bar q^k b)\nonumber\\
&Q_{(quqd)}^{(8)}&&=&& \ (\bar q^j T^At)\epsilon_{jk}(\bar q^k T^A b)\label{loopoperators}
\end{alignat}
where $b$ denotes the right-handed bottom-quark and $\tau^I$ are the Pauli-matrices. This list represents a complete set of four-fermion operators consistent with the assumptions made above and leads to the loop-diagrams shown in Figure \ref{fig:fig3}. The operators $Q_{(uG)}$, $Q_{(quqd)}^{(1)}$ and $Q_{(quqd)}^{(8)}$ are not hermitian. However, only the real parts of their Wilson coefficients enter the final expression for the cross section at $\mathcal{O}\left(1/\Lambda^2\right)$. The subtleties regarding minus signs arising from ordering ambiguities in the four-fermion operators are discussed in \cite{Paraskevas:2018mks}.
\FloatBarrier
\subsection{Treatment of $\gamma_5$}
The Dirac structure of the four-fermion operators include chiral projectors which are accompanied by the strictly four-dimensional object $\gamma_5$. There have been many discussions concerning the question of how to regularize divergent amplitudes with a consistent treatment of $\gamma_5$, in particular for triangle diagrams like the ones appearing in this work, see e.g. \cite{Gnendiger:2017pys} for a recent review. After all, a na\"ive application of dimensional regularization (NDR) is at first sight incompatible with a straightforward continuation of $\gamma_5$ whose relation to the other Dirac matrices are unambiguously defined in four dimensions only. In fact, an anti-commuting $\gamma_5$ cannot be extrapolated to arbitrary dimensions together with the trace formula
\begin{align}\label{epsilon}
tr(\gamma^\alpha\gamma^\beta\gamma^\gamma\gamma^\delta\gamma_5)=-4i\epsilon^{\alpha\beta\gamma\delta}
\end{align}
where we have chosen the default sign convention implemented in \textsc{FeynCalc} \cite{Shtabovenko:2020gxv}. However, despite being inconsistent in the first place, NDR is known to lead to correct final results when the potential hard anomalies are separately taken into account as has been reviewed in \cite{Jegerlehner:2000dz}.\\
Irrespective of the particular regularization method, most approaches usually rely on shifting the loop momenta at some point of the calculation, so linearly divergent diagrams are expected to produce non-vanishing surface terms. These in general spoil the underlying gauge invariance and should consequently be adjusted by hand in order to keep gauge invariance intact as was demonstrated in \cite{Zhemchugov:2014dza}. For instance, in the textbook example of a three-point-function of two vector- and one axial-vector current for massless fermions, one can shuffle the anomaly around by virtue of boundary terms in a way that the two vector (gauge) currents are conserved, but the axial current - which in our case corresponds to the four-fermion interaction - is not. Having this example in mind, we fix the superficially divergent boundary terms by requiring manifest gauge invariance at each calculational step. Since in the case at hand the SMEFT operators do not modify the SM gauge couplings within the triangle diagrams (as is the case for example in \cite{Cata:2020crs}), no induced \textit{gauge} anomalies are expected to play a role. Also, SMEFT alone does not make any assumptions about the new physics sector, so anomalies in the latter are beyond the scope of this paper and are therefore neglected. More comments about the four-fermion triangle diagrams can be found in \cite{Degrande:2020evl}. Further discussions about calculational aspects regarding $\gamma^5$ can be found in \cite{Chetyrkin:2012rz,Bednyakov:2012en,Bednyakov:2013cta,Boughezal:2019xpp}.
\subsection{Renormalization}
The self-energy corrections of the quark lines induced by the four-fermion operators do not depend on the momenta and are therefore negligible when on-shell renormalization is performed. In general, however, we also need to renormalize the Wilson coefficients of the SMEFT operators, which is done in the $\overline{MS}$ scheme. \\
Most of the contributions are already finite by themselves and do not need any renormalization. Operator mixing only appears between the operators $Q_{(uG)}$, $Q_{(quqd)}^{(1)}$ and $Q_{(quqd)}^{(8)}$, as the latter two produce UV-divergences in the form of the first one. The tree-level operator $Q_{(uG)}$ is therefore the only one that needs special treatment. To the order under consideration an implicit dependence of its Wilson coefficient on the renormalization scale $\mu$ is introduced, which has to cancel the explicit logarithmic dependence associated with $Q_{(quqd)}^{(1)}$ and $Q_{(quqd)}^{(8)}$.\\
The relevant part of the renormalization group equation is given by \cite{Jenkins:2013wua,Celis:2017hod}
\begin{align}\label{renorm}
\frac{d C_{(uG)}}{d\ln (\mu^2)}=& -\frac{g_sm_b}{16\pi^2\sqrt{2}v}\left( C^ {(1)}_{(quqd)}-\frac{1}{6} C^ {(8)}_{(quqd)}\right)
\end{align}
where $v$ is the vacuum expectation value of the Higgs-field and $C_i$ denotes the respective Wilson coefficient of the operator $Q_i$. When this equation is applied to our total differential cross section, we find that our expressions are indeed independent of the renormalization scale $\mu$. Note that all mixing disappears when the bottom-mass $m_b$ is sent to zero. On the other hand, it provides a useful consistency check when the bottom-mass is fully taken into account.
\section{Numerical results}\label{numerics}
\begin{table}[htbp]
	\centering
	\begin{tabular}{|c|c|c|c|c|c|}
		\hline
		$m_t$ & $m_b$ & $m_H$ & $v$ & $\alpha_s(M_Z)$ & $\Lambda$ \\
		\hline
		$173$ GeV & $4.18$ GeV & $125$ GeV & $246$ GeV & $0.1181$ & $1$ TeV  \\
		\hline
	\end{tabular}
	\caption[Tabelle]{\textit{Values of the input parameters used for the analysis taken from} \cite{Zyla:2020zbs}.}
	\label{tab:tab1}
\end{table}
\noindent This section contains an exploratory analysis of the SMEFT corrections to the SM cross section. The input parameters are given in Table \ref{tab:tab1}. In addition we assume a cut-off scale of $\Lambda= 1$ TeV, which is well beyond the parameters of the SM and set $\mu=m_t$. As QCD corrections at next-to-leading order within the SM (see \cite{Beenakker:1988bq,Nason:1987xz} for the computation) do not change the shape of the total cross section as a function of the center-of-mass energy \cite{Baernreuther:2012}, the overall normalization of our results can possibly be adjusted to fit the more realistic curves. Within the SM the K-factor from next-to-leading order QCD corrections is about $1.5$ \cite{Franzosi:2015osa}, so it seems plausible to apply the same number to the SMEFT process we have in mind. Of course, only a detailed analysis of the relevant QCD corrections could ultimately justify this step. Explicit example calculations for QCD corrections in SMEFT can be found in \cite{Maltoni:2016yxb,Maltoni:2018zvp}. Furthermore, starting from its value at the Z-mass $\alpha_s(M_Z)=0.1181$, we used Version 3 of RunDec \cite{Herren:2017osy} to determine the $\overline{MS}$-value $\alpha_s(1 \ \text{TeV})=0.08916$. The running of the strong coupling constant is also a next-to-leading order QCD effect, so a certain amount of arbitrariness, especially for the total cross section as a function of the center-of-mass energy $\sqrt{s}$ is unavoidable in this context. However, as we intend to only present indicative estimations of the new effects when compared to the SM predictions, this should not impact the overall message.\\
\begin{figure}
	\centering
	\includegraphics[width=0.8\textwidth]{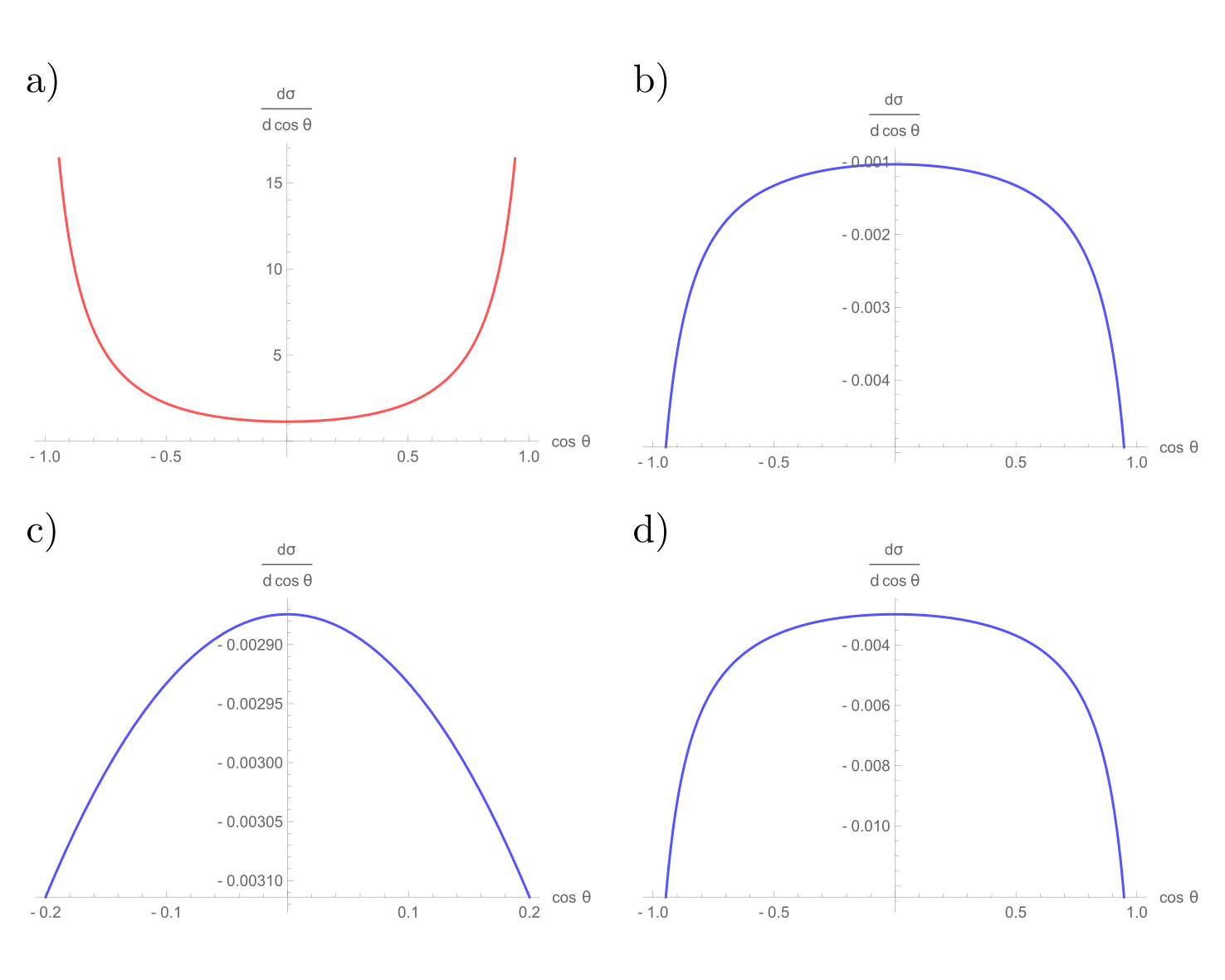}
	\caption{a) \textit{SM differential cross section $(d\sigma/d\cos\theta)_{SM}$.} b - d) \textit{Selected SMEFT corrections to the SM result} b) \textit{$(d\sigma/d\cos\theta)_{Q_{(\varphi G)}}$,} c) \textit{$(d\sigma/d\cos\theta)_{Q_{(u G)}}$ and} d) \textit{$(d\sigma/d\cos\theta)_{Q_{(uu)}}$. The differential cross sections $d\sigma/d\cos\theta$ are given in units of }pb. \textit{The center-of-mass energy was chosen to be $\sqrt{s}=1$} TeV.}
	\label{fig:plot1}
\end{figure}
\begin{figure}
	\centering
	\includegraphics[width=1.0\textwidth]{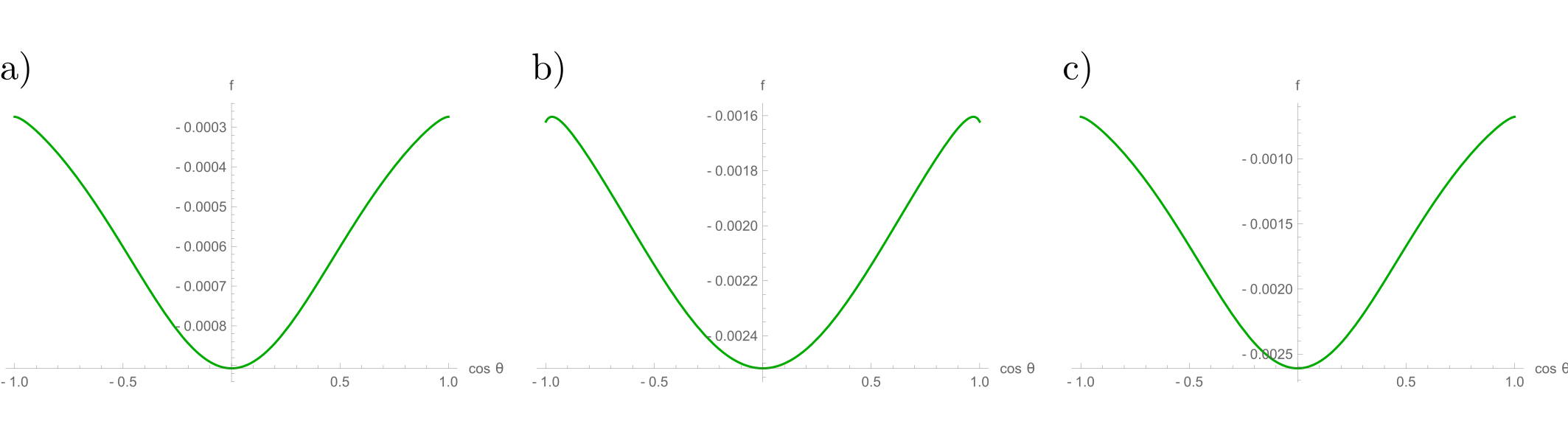}
	\caption{\textit{Selected relative corrections to the SM differential cross section at $\sqrt{s}=1$} TeV \textit{with} a) \textit{$f=(d\sigma/d\cos\theta)_{Q_{(\varphi G)}}/(d\sigma/d\cos\theta)_{SM}$,} b) \textit{$f=(d\sigma/d\cos\theta)_{Q_{(u G)}}/(d\sigma/d\cos\theta)_{SM}$ and} c) \textit{$f=(d\sigma/d\cos\theta)_{Q_{(uu)}}/(d\sigma/d\cos\theta)_{SM}$.}}
	\label{fig:plot2}
\end{figure}
In accordance with our discussion about loop counting above we choose the values of the Wilson coefficients of the operators displayed in Equation (\ref{treeoperators}) to be $1/16\pi^2$, whereas the ones of the four-fermion operators in Equation (\ref{loopoperators}) are set to $1$. Although the positive sign for all coefficients is a mere convention, it does not affect the qualitative outcome of our analysis as we focus only on the magnitudes of the new effects. Of course, despite being well motivated these assumptions may only serve as preliminary estimations for a more complete analysis and should not be taken at face value. Note that within the chosen framework experiments are actually only sensitive to the ratio $C_i/\Lambda^2$. Fixing the cutoff-scale $\Lambda$ and the Wilson coefficients $C_i$ separately therefore corresponds to an artificial split of the actual experimental coefficients.\\
We plot the expected corrections to the SM differential cross section $d\sigma/d\cos\theta$ of the operators $Q_{(\varphi G)}$, $Q_{(uG)}$ and $Q_{(uu)}$ for $\sqrt{s}=1$ TeV in Figures \ref{fig:plot1} and \ref{fig:plot2}. Integrating over the residual angular dependence gives the total cross section $\sigma$ that is plotted against the center-of-mass energy $\sqrt{s}$ in Figure \ref{fig:plot3}. The plots for the remaining operators have similar shapes and are not displayed. In comparison to the SM all SMEFT corrections are suppressed by both $1/\Lambda^2$ and $1/16\pi^2$, so the overall effects are rather small. For the relevant energy regimes, i.e. $\sqrt{s}\approx 1$ TeV, the SMEFT operators give rise to corrections with a relative strength of at best $10^{-3}$ to the SM differential cross section. However, there are some interesting qualitative observations:\\
\\
$\bullet$ Only the operators $Q_{(G)}$, $Q_{(quqd)}^{(1)}$ and $Q_{(qu)}^{(1)}$ have their largest relative impact for high scattering angles. In contrast, all other operators, in particular the ones in Figures \ref{fig:plot1} - \ref{fig:plot3} appear to be more significant for low scattering angles.\\
\\
$\bullet$ The relative correction to the total SM cross section (not plotted) increases most rapidly for the operator $Q_{(qu)}^{(1)}$, reaching the percent level at around $\sqrt{s}=4$ TeV. Of course, possible resonances above $1$ TeV could spoil the validity of the effective theory in this energy regime.\\
\\
$\bullet$ There is a change of sign in the correction of the total cross section just after the threshold energy $\sqrt{s}=2m_t$ for the operators $Q_{(quqd)}^{(1)}$, $Q_{(uu)}$, $Q_{(qq)}^{(1)}$, $Q_{(qq)}^{(3)}$, $Q_{(qu)}^{(1)}$ and $Q_{(qu)}^{(8)}$.\\
\\
The last aspect might in particular be useful for new constraints of the purely right-handed operator $Q_{(uu)}$ (see Figure \ref{fig:plot3} d for the curve). This operator has recently been investigated in \cite{Banelli:2020iau} where an emphasis was put on four-top production at hadron colliders. The combined upper limit from the ATLAS experiment for four-top production is given by $|C_{(uu)}|\le1.9$ for $\Lambda= 1$ TeV \cite{Aaboud:2018jsj}. This does not represent a significant enhancement for the strength of the overall impact. In fact, hypothetical strong couplings $\sim 4\pi$ to the new physics in the top-sector would lead us to superficially expect a numerical value around $|C_{(uu)}|\approx16\pi^2$ if - na\"ively trusting perturbation theory - the four-fermion operator were generated by new exchange processes. The smallness of the actual value, however, indicates significant limitations for such strong coupling scenarios and thus also for the values of $C_{(G)}$, $C_{(\varphi G)}$ and $C_{(uG)}$ (see below).\\
Apart from the top-sector, the Higgs-sector is predestined for strong couplings in the new physics domain that could enhance certain Wilson coefficients involving the Higgs-field. Scenarios with strong dynamics of electroweak symmetry breaking can be described in full generality by the Higgs-Electroweak Chiral Lagrangian \cite{Buchalla:2013rka,Buchalla:2016bse}. In particular, it was pointed out that matching the latter to SMEFT generates the operator class of $Q_{(\varphi G)}$ without the extra loop factor $1/16\pi^2$, in which case its Wilson coefficients should be treated as $\mathcal{O}(1)$-numbers. Indeed, keeping track of all possible weak coupling constants for a given operator generated in a strongly coupled Higgs scenario reveals that the chiral dimensions, i.e. the loop order is effectively reduced by the Higgs-field. The operator $Q_{(\varphi G)}$ then represents the dominant SMEFT correction to gluon fusion top-pair production with relative impact to the SM differential cross section in the low percent range. In contrast, the operator $Q_{(u G)}$ can still only be generated at next-to-leading order, so here our discussion above remains valid even for strong couplings in the Higgs-sector. See also \cite{Buchalla:2018yce} and the comments in Chapter 2.1 in \cite{DiMicco:2019ngk} for gluon fusion Higgs-pair production. Within the Higgs-Electroweak Chiral Lagrangian, the gluon-gluon-Higgs coupling is parameterized by a coefficient $c_g$, that can directly be translated to $C_{(\varphi G)}$ via the formula
\begin{align}
C_{(\varphi G)}=\frac{\Lambda^2\alpha_s}{8\pi v^2}c_g\sim 0.08 \ c_g
\end{align}
where the parameters are defined in Table \ref{tab:tab1}. Keep in mind, that in a strongly coupled scenario, there is no decoupling of the effective theory from its cut-off scale $\Lambda_{s.c.}$, as it is related to the low energy scale $v$ by $\Lambda_{s.c.}=4\pi v$. A loop factor $1/16\pi^2$ can therefore be traded against the expansion parameter $v^2/\Lambda_{s.c.}^2$ and vice versa. The experimental value for $c_g$ can be found in \cite{deBlas:2018tjm} and is approximately given by $c_g=-0.01\pm0.08$, indicating that the overall effects are still rather small.\\
Experimental constraints for the remaining tree-level operators can be found in \cite{Krauss:2016ely,Hirschi:2018etq,Brivio:2019ius,Buckley:2015lku,AguilarSaavedra:2018nen,Hartland:2019bjb} and are at best given by $|C_{(uG)}|\le 0.78$ and $|C_{(G)}|\le 0.037$ for $\Lambda=1$ TeV, depending on the fitting procedure. While the latter value seems more plausible in light of our assumptions, the experimental bounds of the former are not as constraining. As a matter of fact, both numbers are still well above their natural value of around $1/16\pi^2\approx 0.0063$.
\newpage
\section{Conclusion and Outlook}
In this paper we have computed the differential cross section for gluon fusion top-pair production in SMEFT including single insertions of operators of canonical dimension six. Systematic power counting rules relying on realistic assumptions about the new physics sector lead us to consider the tree-level contributions on the same footing as one-loop contributions arising from four-fermion operators. Our calculation serves as an example of how to treat operators featuring gluon field strength tensors in a consistent manner within the perturbative expansion. In particular, it illustrates how this class needs to be accompanied by four-fermion operators with explicit loop suppression to ensure working with a complete set of operators for a given loop order. As a result, the overall SMEFT effects are rather small. Meanwhile, since next-to-leading order QCD corrections are expected to be of great importance to the process under concern, a broader analysis should include them as well. We postpone a general phenomenological discussion together with a more complete analysis concerning the parameter space of the Wilson coefficients to future works.
\begin{figure}
	\centering
	\includegraphics[width=0.9\textwidth]{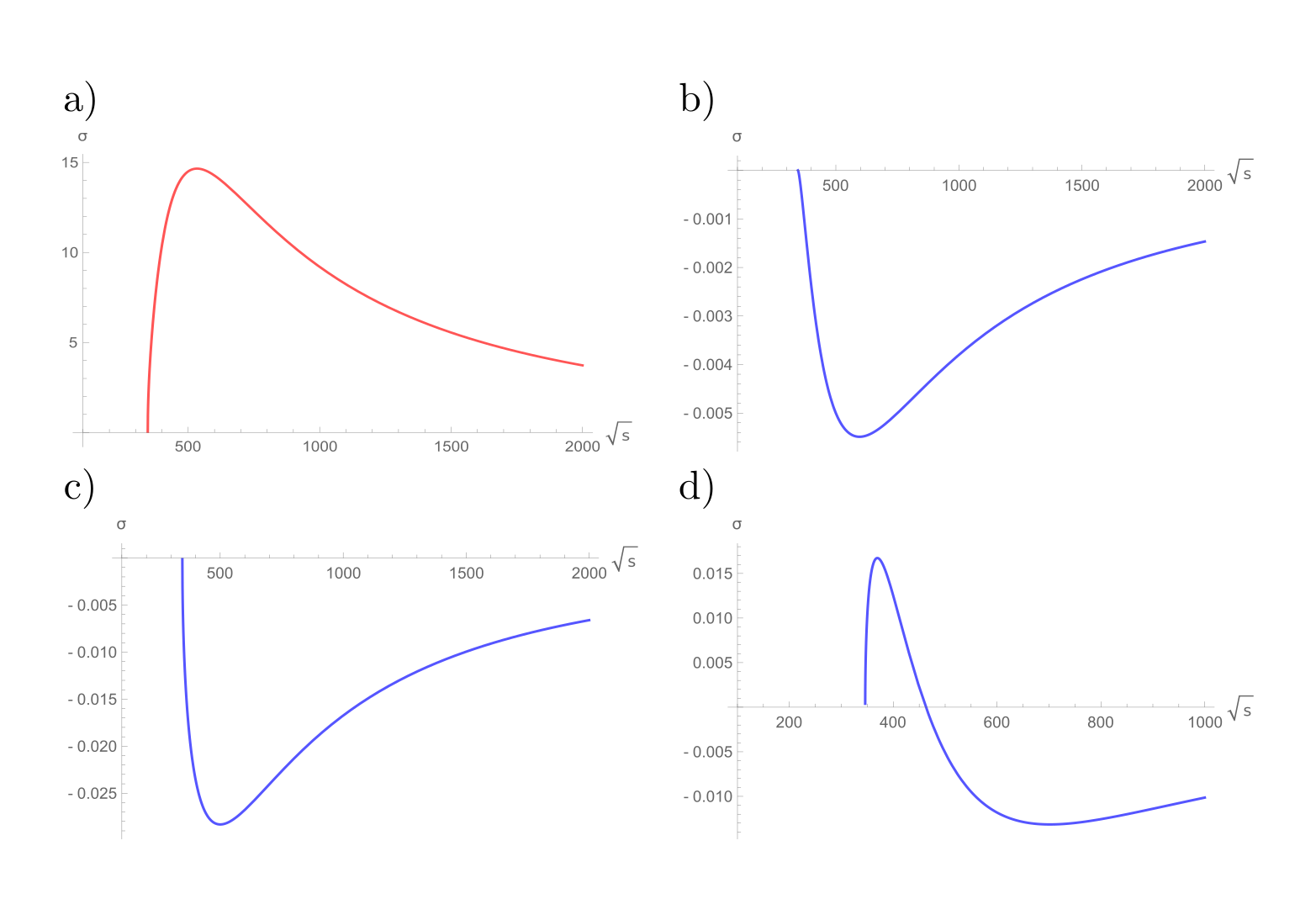}
	\caption{a) \textit{SM total cross section $\sigma_{SM}$.} b - d) \textit{Selected SMEFT corrections to the SM result} b) \textit{$\sigma_{Q_{(\varphi G)}}$,} c) \textit{$\sigma_{Q_{(u G)}}$ and} d) \textit{$\sigma_{Q_{(uu)}}$. The center-of-mass energies $\sqrt{s}$ are given in units of} GeV,\textit{ whereas the total cross sections $\sigma$ are again given in units of} pb.}
	\label{fig:plot3}
\end{figure}
\FloatBarrier
\section*{Acknowledgments}
I would like to thank Gerhard Buchalla for useful discussions at the different stages of
this work and valuable comments on the manuscript. I am supported by the Studienstiftung des Deutschen Volkes and by
the Deutsche Forschungsgemeinschaft (DFG, German Research
Foundation) under grant BU 1391/2-2 (project number 261324988) and by the DFG under Germany’s Excellence
Strategy – EXC-2094 – 390783311 ”ORIGINS”.
\appendix
\section{Corrections to the SM differential cross section}	\label{appA}
This appendix features a complete list of the analytic expressions for the corrections to the SM differential cross section due to the SMEFT operators. The differential cross section including the SMEFT corrections can be written as
 \begin{align}
\left(\frac{d\sigma}{d\Omega}\right)_{SM}+\left(\frac{d\sigma}{d\Omega}\right)_{SMEFT}
\end{align}
where $\left(d\sigma/d\Omega\right)_{SMEFT}$ denotes the sum of all contributions of the dimension-six operators under concern.\\
The relevant real parts of the Feynman parameter integrals are given by
\begin{align}
&S_{1}(a_i)=\mathrm{Re}\Biggl\{\int_0^1 d z\int_0^{1-z} dy \frac{1}{1-4yza_i-i\eta}\Biggr\}=\frac{-\ln^2(4a_i)+\pi^2}{8a_i}+\mathcal{O}\left(\frac{1}{a_i^2}\right) \\
&S_{2}(a_i)=\mathrm{Re}\Biggl\{\int_0^1 d z\int_0^{1-z} dy \frac{y}{1-4yza_i-i\eta}\Biggr\}=\frac{-\ln(4a_i)+2}{4a_i}+\mathcal{O}\left(\frac{1}{a_i^2}\right)
\end{align}
where $a_i=s/4m_i^2$ for the top- and the bottom-mass and $\eta$ is a small positive number and a center-of-mass energy above the threshold for top-pair production is assumed.\\
Defining $\widetilde C_i=C_i/\Lambda^2$ and $\beta=\sqrt{1-4m_t^2/s}$, the single analytic expressions for the contributions of the SMEFT operators are given by the following list:\\
 \begin{alignat}{3}
&\left(\frac{d\sigma}{d\Omega}\right)_{Q_{(G)}}=&&-\widetilde C_{(G)}\frac{\alpha_s^2m_t^2\beta}{32g_ss}\frac{9  (t-u)^2}{(m_t^2-t) (m_t^2-u)}\\
&\left(\frac{d\sigma}{d\Omega}\right)_{Q_{(\varphi G)}}=&& \ \ \widetilde C_{(\varphi G)}\frac{\alpha_sm_t^2\beta^3}{64\pi}\frac{ s^2}{(m_t^2-t) (m_t^2-u) (m_h^2-s)}\\
&\left(\frac{d\sigma}{d\Omega}\right)_{Q_{(uG)}}=&&-\mathrm{Re} \{\widetilde C_{(uG) }(\mu)\}\frac{\alpha_s^{3/2}vm_t\beta}{24\sqrt{2\pi}s}\frac {7 m_t^4 - 7 m_t^2 (t + u) + 4 t^2 - 
	t u + 4 u^2}{(m_t^2-t) (m_t^2-u)}
\end{alignat}
\newpage
\begin{alignat}{3}
&\left(\frac{d\sigma}{d\Omega}\right)_{Q^{(1)}_{(quqd)}}=&&-\frac{\mathrm{Re} \{\widetilde C^{(1)}_{(quqd)}\}}{16\pi^2}\frac{\alpha_s^2m_t\beta}{192m_bs}\frac {1}{  (m_t^2-t) (m_t^2-u)}\cdot\nonumber\\
&&&\cdot\biggl(19 s^2 \Bigl(2m_t^2 S_{1}(a_b)+m_b^2\beta^2\bigl(2S_{1}(a_b)-1\bigr)\Bigr)+\nonumber\\
&&&+8  m_b^2 \bigl(7 m_t^4-7 m_t^2 (t+u)+4 t^2-t u+4 u^2\bigr)\ln\frac{\mu^2}{m_b^2}\biggr) \\
&\left(\frac{d\sigma}{d\Omega}\right)_{Q^{(8)}_{(quqd)}}=&&-\frac{\mathrm{Re} \{\widetilde C^{(8)}_{(quqd)}\}}{16\pi^2}\frac{\alpha_s^2m_t\beta}{1152m_bs}\frac {1}{  (m_t^2-t) (m_t^2-u)}\cdot\nonumber\\
&&&\cdot\biggl(41 s^2 \Bigl(2m_t^2 S_{1}(a_b)+m_b^2\beta^2\bigl(2S_{1}(a_b)-1\bigr)\Bigr)+\nonumber\\
&&&-8  m_b^2 \bigl(7 m_t^4-7 m_t^2 (t+u)+4 t^2-t u+4 u^2\bigr)\ln\frac{\mu^2}{m_b^2}\biggr)\\
&\left(\frac{d\sigma}{d\Omega}\right)_{Q_{(qd)}^{(1)}}=&&-\frac{\widetilde C_{(qd)}^{(1)}}{16\pi^2}\frac{\alpha_s^2 m_t^2 s\beta}{32}\frac{  2S_{1}(a_b)-1}{ (m_t^2-t) (m_t^2-u)}\\
&\left(\frac{d\sigma}{d\Omega}\right)_{Q_{(ud)}^{(1)}}=&&\frac{\widetilde C_{(ud)}^{(1)}}{16\pi^2}\frac{\alpha_s^2m_t^2 s\beta}{32}\frac{ 2S_{1}(a_b)-1}{(m_t^2-t) (m_t^2-u)}\\
&\left(\frac{d\sigma}{d\Omega}\right)_{Q^{(8)}_{(qd)}}=&&-\frac{\widetilde C_{(qd)}^{(8)}}{16\pi^2}\frac{\alpha_s^2m_t^2\beta}{384s}\frac{ 5s^2\bigl( 2S_{1}(a_b)-1\bigr)+3(t-u)^2\bigl( 6S_{1}(a_b)-12S_{2}(a_b)-1\bigr)}{ (m_t^2-t) (m_t^2-u)}\\
&\left(\frac{d\sigma}{d\Omega}\right)_{Q^{(8)}_{(ud)}}=&&\frac{\widetilde C_{(ud)}^{(8)}}{16\pi^2}\frac{\alpha_s^2m_t^2\beta}{384s}\frac{5s^2\bigl( 2S_{1}(a_b)-1\bigr)-3(t-u)^2\bigl( 6S_{1}(a_b)-12S_{2}(a_b)-1\bigr)}{(m_t^2-t) (m_t^2-u)}\\
&\left(\frac{d\sigma}{d\Omega}\right)_{Q_{(uu)}}=&&\frac{\widetilde C_{(uu)}}{16\pi^2}\frac{\alpha_s^2m_t^2\beta}{96s}\frac{13s^2\bigl( 2S_{1}(a_t)-1\bigr)-3(t-u)^2\bigl( 6S_{1}(a_t)-12S_{2}(a_t)-1\bigr)}{ (m_t^2-t) (m_t^2-u)}\\
&\left(\frac{d\sigma}{d\Omega}\right)_{Q^{(1)}_{(qq)}}=&&\frac{\widetilde C^{(1)}_{(qq)}}{16\pi^2}\frac{\alpha_s^2m_t^2\beta}{96s}\frac {1}{ (m_t^2-t) (m_t^2-u)}\Bigl(6s^2\bigl( 2S_{1}(a_b)-1\bigr)+\nonumber\\
&&&+13s^2\bigl( 2S_{1}(a_t)-1\bigr)-3(t-u)^2\bigl( 6S_{1}(a_t)-12S_{2}(a_t)-1\bigr)\Bigr)
\end{alignat}
\newpage
\begin{alignat}{3}
&\left(\frac{d\sigma}{d\Omega}\right)_{\widetilde C^{(3)}_{(qq)}}=&& \ \frac{\widetilde C^{(3)}_{(qq)}}{16\pi^2}\frac{\alpha_s^2m_t^2\beta}{96s}\frac{1}{ (m_t^2-t) (m_t^2-u)}\Bigl(8s^2\bigl( 2S_{1}(a_b)-1\bigr)+\nonumber\\
&&&+13s^2\bigl( 2S_{1}(a_t)-1\bigr)-6(t-u)^2\bigl( 6S_{1}(a_b)-12S_{2}(a_b)-1\bigr)+\nonumber\\
&&&-3(t-u)^2\bigl( 6S_{1}(a_t)-12S_{2}(a_t)-1\bigr)\Bigr)\\
&\left(\frac{d\sigma}{d\Omega}\right)_{Q^{(1)}_{(qu)}}=&&-\frac{\widetilde C^{(1)}_{(qu)}}{16\pi^2}\frac{\alpha_s^2\beta}{96s}\frac{1}{ (m_t^2-t) (m_t^2-u)}\biggl(3 m_t^2 s^2\bigl(2S_{1}(a_b)-1\bigr)+56 m_t^6 +\nonumber\\
&&& +4 m_t^2  \Bigl(14m_t^2s \bigl(2S_{1}(a_t)-1\bigr)-7s^2S_{1}(a_t)+8 t^2-2 t u+8 u^2\Bigr) +\nonumber\\
&&&-8 m_t^2s^2 \bigl(2S_{1}(a_t)-1\bigr)-56 m_t^4  (t+u) +14 s^3 S_{1}(a_t) \biggr)\\
&\left(\frac{d\sigma}{d\Omega}\right)_{Q^{(8)}_{(qu)}}=&&-\frac{\widetilde C^{(8)}_{(qu)}}{16\pi^2}\frac{\alpha_s^2\beta}{1152s}\frac {1}{ (m_t^2-t) (m_t^2-u)}\biggl(15 m_t^2 s^2\bigl(2S_{1}(a_b)-1\bigr) +\nonumber\\
&&&-14 m_t^2 s^2\bigl(2S_{1}(a_t)-1\bigr) -112 m_t^6 +112 m_t^4  (t+u)+\nonumber\\
&&&-8 m_t^2  \Bigl(-22m_t^2s \bigl(2S_{1}(a_t)-1\bigr) +11s^2 S_{1}(a_t) +8 t^2-2 t u+8 u^2\Bigr)+\nonumber\\
&&&+9 m_t^2(t-u)^2\bigl(6S_{1}(a_b)-12S_{2}(a_b)-1\bigr) +44 s^3 S_{1}(a_t) \biggr)
\end{alignat}
\newpage
	\bibliographystyle{utphys}
\bibliography{references}
\end{document}